\documentclass[letterpaper]{article} 
\usepackage{aaai25}  
\usepackage{times}  
\usepackage{helvet}  
\usepackage{courier}  
\usepackage[hyphens]{url}  
\usepackage{graphicx} 
\usepackage{natbib}  
\usepackage{caption} 
\usepackage{algorithm}
\usepackage{algorithmic}
\usepackage{newfloat}
\usepackage{listings}
\DeclareCaptionStyle{ruled}{labelfont=normalfont,labelsep=colon,strut=off} 
\floatstyle{ruled}
\newfloat{listing}{tb}{lst}{}
\floatname{listing}{Listing}
\usepackage{makecell}

\usepackage{booktabs} 
\usepackage{graphicx}
\usepackage{subfigure}
\usepackage{epstopdf}
\usepackage{epsfig}
\usepackage{xspace}
\usepackage{amssymb,amsmath}
\usepackage{enumitem}
\usepackage{algorithm}
\usepackage{algorithmic}
\usepackage{amsthm}
\usepackage{caption}
\usepackage{multirow}
\usepackage{threeparttable}
\usepackage{diagbox}

\usepackage{url}

\usepackage{url}

\newcommand{\ie}{\textit{i}.\textit{e}.}

\newtheorem{Def}{Definition}

\newtheorem*{Pro*}{Problem}

\newcommand{\model}{MT-Link\xspace}

\title{Correlation-Attention Masked Temporal Transformer for User Identity Linkage Using Heterogeneous Mobility Data}

\author{
    Ziang Yan\textsuperscript{\rm 1},
    Xingyu Zhao\textsuperscript{\rm 1}, 
    Hanqing Ma\textsuperscript{\rm 1}, \\
    Wei Chen\textsuperscript{\rm 2},
    Jianpeng Qi\textsuperscript{\rm 1},
    Yanwei Yu\textsuperscript{\rm 1}\thanks{Corresponding author: Yanwei Yu.},
    Junyu Dong\textsuperscript{\rm 1}
}
\affiliations{
    \textsuperscript{\rm 1}Faculty of Information Science and Engineering, Ocean University of China, Qingdao, China\\
    \textsuperscript{\rm 2}The Hong Kong University of Science and Technology (Guangzhou)\\
    
    \{yza, zhaoxingyu, mahanqing\}@stu.ouc.edu.cn, onedeanxxx@gmail.com,\\ \{qijianpeng, yuyanwei, dongjunyu\}@ouc.edu.cn
}

\usepackage{bibentry}

\begin{document}

\maketitle

\begin{abstract}
With the rise of social media and Location-Based Social Networks (LBSN), check-in data across platforms has become crucial for User Identity Linkage (UIL). These data not only reveal users' spatio-temporal information but also provide insights into their behavior patterns and interests. However, cross-platform identity linkage faces challenges like poor data quality, high sparsity, and noise interference, which hinder existing methods from extracting cross-platform user information. To address these issues, we propose a Correlation-Attention \textbf{\underline{M}}asked \textbf{\underline{T}}ransformer for
User Identity \textbf{\underline{Link}}age Network (\model), a transformer-based framework to enhance model performance by learning spatio-temporal co-occurrence patterns of cross-platform users. Our model effectively captures spatio-temporal co-occurrence in cross-platform user check-in sequences. It employs a correlation attention mechanism to detect the spatio-temporal co-occurrence between user check-in sequences. Guided by attention weight maps, the model focuses on co-occurrence points while filtering out noise, ultimately improving classification performance. Experimental results show that our model significantly outperforms state-of-the-art baselines by 12.92\%$\sim$17.76\% and 5.80\%$\sim$8.38\% improvements in terms of Macro-F1 and Area Under Curve (AUC). 
\begin{links}
\link{Code}{https://github.com/DrivenA/MT-Link}
\end{links}
\end{abstract}

\section{Introduction}
Location-Based Social Network (LBSN) services like Twitter and Foursquare have made daily life more convenient, generating vast amounts of spatio-temporal human mobility data~\cite{qin2023graph}, such as point-of-interest (POI) check-in sequences~\cite{zhao2020go}.  The availability of such spatio-temporal data provides a foundation for exploring User Identity Linkage (UIL) tasks~\cite{riederer2016linking}, which holds significant potential in areas such as recommendation systems~\cite{chen2020multi}, user behavior analysis, and privacy protection~\cite{qi2018two}. Several studies~\cite{goga2013exploiting, rossi2014s, naini2015you} have leveraged spatio-temporal check-in data to match user identities based on mobility trajectories, demonstrating the effectiveness of such data.

Previous methods have achieved some success~\cite{goga2013exploiting, rossi2014s, naini2015you}, but most are limited by the discrete nature of the data provided by the datasets. The performance of these methods is often constrained by the dataset distribution and lacks a deep exploration of hidden patterns within user check-in sequences. Despite these limitations, data mining approaches have made significant progress in addressing the spatio-temporal UIL problem~\cite{riederer2016linking, basik2017spatio, li2019lisc, basik2020slim}. These methods effectively link the same user across different platforms by focusing on spatial and temporal correlations. Recently, clustering-based methods ~\cite{ding2020user, ma2022cp, chen2017exploiting, xue2021kmul} and kernel density estimation methods~\cite{chen2018effective, chen2023hful} have further attempted to solve the challenges posed by sparse spatio-temporal check-in data, thereby improving identification accuracy. Notably, deep learning approaches~\cite{feng2019dplink, feng2020user, LI2023357} have also achieved impressive success in capturing abstract representations of user check-in sequences, demonstrating significant potential in modeling complex spatio-temporal data.

\begin{figure}
    \begin{center}
    \includegraphics[width=0.47\textwidth]{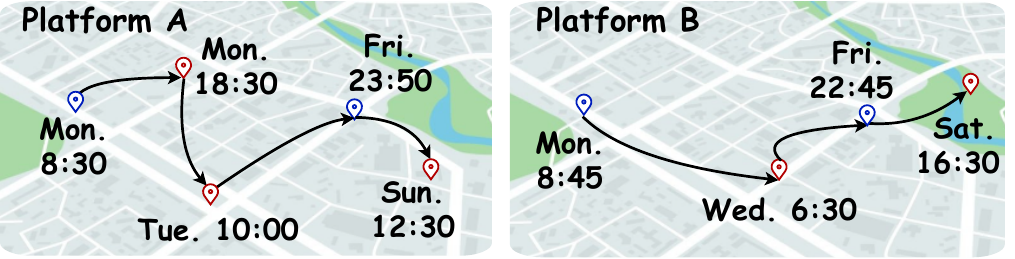}
    \vspace{-5mm}
    \caption{Check-ins of the same user on different platforms.}
    \vspace{-8mm}
    \label{fig:intro}
    \end{center}
\end{figure}

Although significant progress has been made in spatio-temporal user identity linkage, we identify three key limitations in most existing methods. \textit{First, current research mainly focuses on solving the problem of data sparsity, but introduces inherent noise in the check-in sequence and ignores the information of co-occurrence points}. For example, 
Figure~\ref{fig:intro} clearly shows that the check-in frequency of the same user varies across different platforms. We define check-ins that occur at the same time and place as spatio-temporal co-occurrence points. If we consider all check-in points to determine whether these check-in points are generated by the same user, random check-ins (\ie, the red points) may interfere with the judgment, increasing the likelihood of errors. However, by focusing on co-occurrence points, we can clearly see that platform A and platform B share the same check-in patterns at specific times and locations (\ie, the blue points), and cross-platform user co-occurrence information is easier to mine.
\textit{Second, current deep learning approaches typically model each user's spatio-temporal sequence independently, without simultaneously considering the spatio-temporal co-occurrence of key points across different user trajectories}. For example, in the different trajectories on A and B shown in Figure 1, if we independently learn the specific patterns of these trajectories, noise interference and the lack of co-occurrence capture at key points could negatively impact the model’s classification performance, leading to incorrect associations between the current user and unrelated users. \textit{Third, the process of manually constructing features and frequent pattern mining of existing data mining methods is not only very time-consuming, but also fails to effectively utilize known real labels in the data.} Lack of true value labels leads to the inability to learn the relationship between input sample pairs and users, which affects the performance of the model.

To address these challenges, we propose a transformer-based symmetric deep learning model called Correlation Attention \textbf{\underline{M}}asked \textbf{\underline{T}}ransformer for
User Identity \textbf{\underline{Link}}age Network (\model). Specifically, we design a temporal transformer encoder and a masked transformer encoder. The dense representations from the spatio-temporal embedding layer are fed into the temporal transformer encoder to model the check-in sequences. Then, we use a correlation attention block to collaboratively capture the spatio-temporal co-occurrences between the check-in sequence representations of users on different platforms. Based on the captured spatio-temporal co-occurrence, we guide the masking of the tokens with the lowest attention in the dense representations, helping us to filter out noise points and retain co-occurrence points.  Finally, the user identity linkage layer links the outputs of the user check-in sequences from both platforms and produces the prediction probability. The results on four real-world cross-platform datasets demonstrate that our model outperforms the latest deep learning methods and data mining approaches in the spatio-temporal UIL task. In summary, our contributions are as follows:
\begin{itemize}
    \item We propose a novel \model method that combines a masking mechanism with a masked transformer encoder to retain key information and ignore noise, enhancing the co-occurrence between information,  thereby improving user identity linkage performance.
    \item We propose the correlation attention mechanism to capture spatio-temporal co-occurrences between cross-platform user check-in sequence representations, better guiding our masking process.
    \item We conduct extensive experimental evaluations on four cross-platform datasets. Experimental results show that our model significantly outperforms state-of-the-art baselines by 12.92\%$\sim$17.76\% and 5.80\%$\sim$8.38\% improvements in terms of Macro-F1 and Area Under Curve (AUC).
\end{itemize} 
\section{Related Work}


The existing research on UIL can be roughly divided into two categories: UIL based on traditional data mining methods and UIL based on deep learning methods.

\textbf{Data Mining-Based Methods.}
In research based on traditional data mining methods, 
\cite{naini2015you} focuses on calculating the frequency of a user's visits to each location, then uses Kullback-Leibler divergence to define the similarity score between two histograms. \cite{riederer2016linking} is an alignment algorithm that calculates an affinity score based on timestamped location data, and then uses a maximum weighted matching scheme to identify the most likely matching user identities. STUL~\cite{chen2017exploiting} extracts spatio-temporal features using density clustering and Gaussian mixture models to measure user similarity. Subsequently, 
~\cite{chen2018effective} employs kernel density estimation to improve accuracy by mitigating sparsity issues. ~\cite{wang2018you} uses a set matching algorithm to identify candidate sets of the same user, employing Bayesian inference for ranking confidence scores to determine cross-platform data linkage. CP-Link~\cite{ding2020user} and its extensions~\cite{ma2022cp} utilize behavior patterns to mine frequent locations and apply an improved dynamic time-warping method for similarity calculation.

\textbf{Deep Learning-Based Methods.} 
To overcome the feature difficulties caused by the high heterogeneity of mobile data from different sources, the deep learning framework DPLink for UIL is first proposed. DPLink~\cite{feng2019dplink} and its extensions~\cite{feng2020user} address high data heterogeneity across platforms with a pre-training strategy, pioneering the use of end-to-end deep learning for UIL. Subsequently, \cite{10368338} proposes a graph convolutional network method to learn user representations from location-based social relationships, aggregating information from surrounding nodes and generating embeddings that comprehensively represent users and locations. \cite{101007} proposes using graph convolution aggregation to supplement the missing neighborhoods of nodes in two co-author networks, where the deviation of adjacent nodes is trained from two well-structured head nodes and corrected locally.

\textit{Most existing UIL methods rely on data mining techniques, which have limitations in capturing deep co-occurrence across platforms, particularly with nonlinear, heterogeneous, and complex data. Traditional deep learning approaches also fall short in leveraging advanced models to capture the intrinsic spatio-temporal co-occurrences in cross-platform check-in sequences.}

\section{Problem Definition}
\label{sec:problem}

\begin{figure*}
    \begin{center}
    \includegraphics[width=1.0\textwidth]{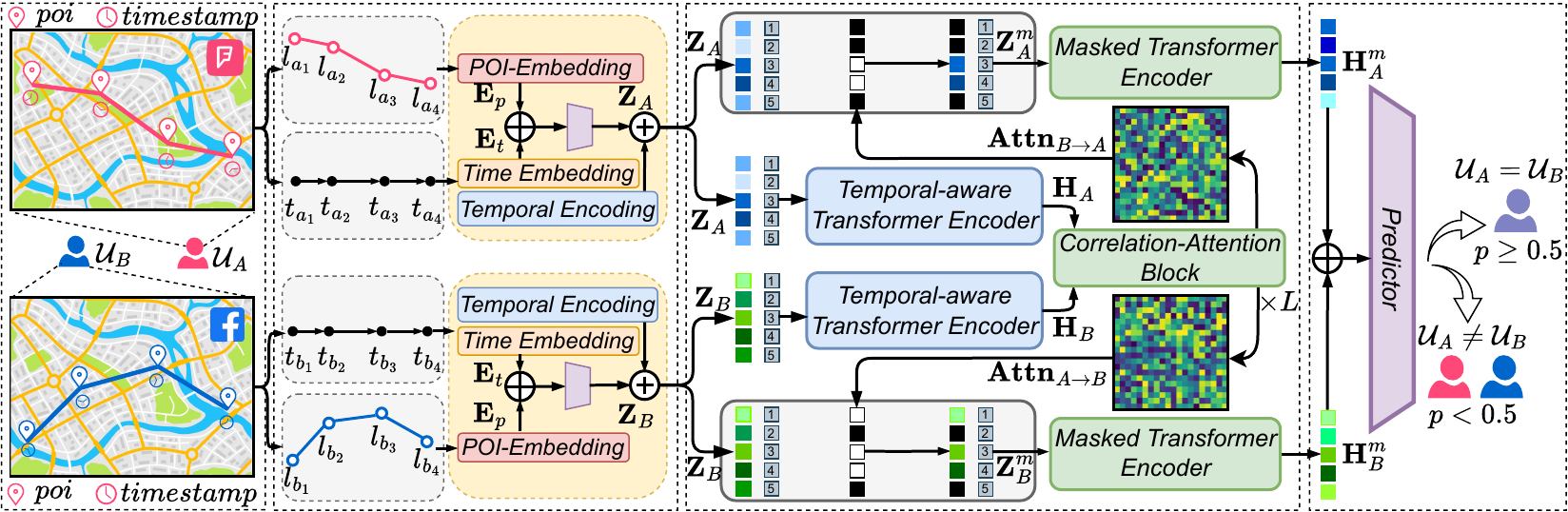}
    \vspace{-5mm}
    \caption{The overview of the proposed framework.}
    \vspace{-5mm}
    \label{fig:overview}
    \end{center}
\end{figure*}

We define the user sets ${\cal U}_A = \{ u_1,u_2,\dots,u_m\}$ and ${\cal U}_B = \{ u_1,u_2,\dots,u_n\}$ as the collections of registered users on two different social platforms $A$ and $B$, where $m$ and $n$ are the number of users in ${\cal U}_A$ and ${\cal U}_B$, respectively.

\begin{Def}[Check-in Point]
    A check-in point is represented by a triplet $(u,t,p)$. Here, $u$ is the user ID on the current social platform, $t$ is the timestamp of the check-in, and $p$ is the POI. 
\end{Def} 

\begin{Def}[Check-in Sequence]
    For $\forall u_i \in \cal U$ in $A$ and $B$, all check-in points are sorted in chronological order by $t$, forming a long spatial temporal sequence ${\cal T}_{u_i} = \{(u_i,t_1,p_1), (u_i,t_2,p_2), \ldots, (u_i,t_k,p_k)\}$, where $i$ is the $i$-th user and $k$ represents the numbers of triplet in the sequence.
\end{Def} 

\begin{Pro*}[User Identity Linkage]
\label{def:prob}
    $\forall u^{A}_i \in {\cal U}_A$ and $\forall u^{B}_j \in {\cal U}_B$, they have their respective check-in sequences ${\cal T}^{A}_{u_i}$ and ${\cal T}^{B}_{u_j}$. The goal is to construct a mapping function $f_\theta(\cdot)$ for these check-in sequences, 
    \begin{equation}
    \label{eq.one}
        {f}({\cal T}_{{u_i}}^A,{{\cal T}_{{u_j}}^B} ;\theta){\rm{ = }}\left\{ {
        \begin{array}{*{20}{c}}
        {{\rm{1,}}\quad u_i^A,u_j^B{\rm{ \:is \:the \:same \:user}}{\rm{.}}}\\
        {{\rm{0,}}\quad u_i^A,u_j^B{\rm{ \:is \:different \:user}}{\rm{.}}}
        \end{array}} \right.
    \end{equation}
    where $\theta$ is the parameter of the function. 
\end{Pro*} 
\section{Methodology}

The architecture of the \model network is depicted in Figure~\ref{fig:overview}.
The primary components of the network include: (1) \textit{Spatial-Temporal Embedding Layer}, (2)\textit{Temporal Transformer Encoder}, (3) \textit{Correlation Attention Block}, (4) \textit{Masked Transformer Encoder}, and (5) \textit{User Identity Linkage Layer}. The following sections will introduce each module in detail.

\subsection{Spatial-Temporal Embedding Layer} 
Discrete data are typically high-dimensional and sparse, leading to the curse of dimensionality, which complicates processing and analysis~\cite{berisha2021digital}. Traditional one-hot encoding is unsuitable for our task due to its inability to capture spatio-temporal semantics and its high computational cost~\cite{rodriguez2018beyond}.
To better capture user trajectory semantics across platforms, we model check-in sequences ${{\cal T}}$ from both spatial and temporal dimensions. Specifically, each POI in the check-in sequence ${{\cal T}}$, \ie, POI $p$, is represented by $|\cal P|$-dimensional one-hot vector. The time of the check-in sequence, \ie, timestamp $t$, is represented by $|T|$-dimensional one-hot vector by day of the month. By learning the embeddings of each POI, denoted by ${{\bf{E}}_p} \in {{\mathbb R}^{\left| \cal P \right| \times {d_p}}}$ and time slot denoted by ${{\bf{E}}_t} \in {{\mathbb R}^{\left| T \right| \times {d_t}}}$, we can capture the semantic information of POIs within each time slot and the relationships between different time slots in the check-in sequence ${{\cal T}}$. Finally, we obtain the embedded sequence ${\bf{X}} = f_{st}({{\cal T}};{\theta _{st}}) = {(x_1, \ldots ,x_k)}  \in {{\mathbb R}^{k \times d}}$, where $f_{st}(\cdot)$ denotes the spatial-temporal embedding layer and $\theta_{st}$ is the learnable parameters in the embedding.

\subsection{Temporal Transformer Encoder} 
Transformer model~\cite{vaswani2017attention}, utilizing attention mechanisms, efficiently processes long sequences and overcomes the limitations of traditional RNNs. In our task, it is crucial to model long spatio-temporal sequences while preserving temporal correlations to achieve high-quality intra-sequence and inter-sequence representations. Inspired by ~\cite{chen2022mutual}, we replace the original transformer's positional encoding with a temporal positional encoding, allowing time slot information to reflect the temporal changes in the check-in sequences ${{\cal T}}$:
\begin{equation}
{[{\mathop{\rm TE}\nolimits} ({t_i})]_j} = \frac{1}{{\sqrt {{d}} }} \cdot \cos ({{\bf{w}}_j} \cdot {t_i} + {{\bf{b}}_j}),
\end{equation}
where ${w_j}$ and ${b_j}$ are learnable parameters, $j$ is the $j$-th order in the dimension of embedded sequence($j \le d$). $i$ is the $i$-th check-in point in check-in sequence ${\cal T}$. For any two check-in points in ${\cal T}$, their relative visited check-in time information can be represented as:
\begin{equation}
{\mathop{\rm TE}\nolimits} ({t_i}){\mathop{\rm TE}\nolimits} {({t_i} + {\Delta _t})^{\bf{T}}} = \sum\limits_{j = 1}^{{d}} {\cos ({{\bf{w}}_j}{\Delta _t}).} 
\end{equation}

We obtain the final embedded representations for ${\cal T}$ by ${{\bf{z}}_i} = {{\bf{x}}_i} + {\mathop{\rm TE}\nolimits} ({t_i})$ with their corresponding temporal positional encoding ${\rm TE}(\cdot)$, resulting in ${\bf{Z}} = [{{\bf{z}}_1}, \ldots ,{{\bf{z}}_k}] \in {{\mathbb R}^{k \times d}}$. We denote our temporal transformer encoder as ${f_{\rm T}}(\cdot)$ and feed the embedded representations ${\bf{Z}}$ into it for encoding. Through this process, we obtain the spatio-temporal context representation vectors ${\bf{H}} \in {{\mathbb R}^{k \times d}}$ for the user check-in sequences on a single platform.

\subsection{Correlation Attention Block} 
To capture the similar semantic information of check-in sequences from platforms $A$ and $B$, we introduce the Correlation Attention Block of $L$ layers to capture the co-occurrence of user check-in sequences. Similar methods have been shown to effectively capture cross-domain relationships by leveraging attention mechanisms~\cite{wang2024timexer}, improving the model's ability to understand and link data across different domains.

In our bidirectional symmetric structure for platform $A$, we obtain the high-level sequence representation ${\bf{H}}_A$ by encoding the previous step ${\bf{Z}}_A$ through the temporal transformer ${f_{\rm T}}(\cdot)$ from platform $A$ as the query vector. \ie, ${\bf{Q}}_A\in {{\mathbb R}^{k \times d}}$ after linear transformation and the high-level sequence representation ${\bf{H}}_B$ by encoding the previous step's ${\bf{Z}}_B$ through the temporal transformer ${f_{\rm T}}(\cdot)$ from platform $B$ as the key and value vectors after linear transformation.\ie, ${\bf{K}}_B\in {{\mathbb R}^{k \times d}}$, ${\bf{V}}_B\in {{\mathbb R}^{k \times d}}$. We then concatenate the output of the attention layer and the high-level sequence representation from platform $A$ as the input of the next attention layer, and this process is repeated. For platform $B$, the logic is identical to that of $A$. This process can be represented as follows:
\begin{equation}
\begin{split}
{\bf{Q}}_A^{l + 1} = {\mathop{\rm LN}\nolimits} \left( {{\bf{\hat Q}}_A^l + {\mathop{\rm CrossAttn}\nolimits} ({\bf{\hat Q}}_A^l,{{\bf{K}}_B},{{\bf{V}}_B})} \right),
\\
{\bf{Q}}_B^{l + 1} = {\mathop{\rm LN}\nolimits} \left( {{\bf{\hat Q}}_B^l + {\mathop{\rm CrossAttn}\nolimits} ({\bf{\hat Q}}_B^l,{{\bf{K}}_A},{{\bf{V}}_A})} \right).
\end{split}
\end{equation}

Here $\mathop{\rm CrossAttn(\cdot)}$ denotes the multi-head cross attention and $l \in L$. ${\bf{\hat Q}}_A^l$ is the $l$-th layer's output and will serve as the input for the next layer. ${\rm LN}(\cdot)$ stands for the Layer Normalization layer.

Finally, we output the attention weight map from the last layer of the block:
\begin{equation}
\begin{split}
{\bf{Attn}}_{A \to B}^{\left| L \right|} = \frac{1}{H}\sum\limits_{h = 1}^H {{\mathop{\rm softmax}\nolimits} ({\bf{Q}}_A^{(\left| L \right|,h)}\frac{{{{({\bf{K}}_B^{(\left| L \right|,h)})}^{\mathop{\rm T}\nolimits} }}}{{\sqrt {{d_k}} }})},
\\
{\bf{Attn}}_{B \to A}^{\left| L \right|} = \frac{1}{H}\sum\limits_{h = 1}^H {{\mathop{\rm softmax}\nolimits} ({\bf{Q}}_B^{(\left| L \right|,h)}\frac{{{{({\bf{K}}_A^{(\left| L \right|,h)})}^{\mathop{\rm T}\nolimits} }}}{{\sqrt {{d_k}} }})},
\end{split}
\end{equation}
where $h$ represents the $h$-th attention head, and we average the attention scores from all $H$ heads to obtain the final output. Each element in ${\bf{Attn}}_{A \to B}\in {{\mathbb R}^{{N_{{Q_A}}} \times {N_{{K_B}}}}}$ represents the relevance score between each position in sequence ${\cal T}_A$ and each position in sequence ${\cal T}_B$. The same logic applies to ${\bf{Attn}}_{B \to A}  \in {{\mathbb R}^{{N_{{Q_B}}} \times {N_{{K_A}}}}}$. 

The collaborative effect of the multi-head self-attention layer in ${f_{\rm T}}(\cdot)$ and the multi-head cross-attention in the correlation attention block allows for a more comprehensive representation of both intra-sequence and inter-sequence correlations within the check-in sequences ${\cal T}_A$ and ${\cal T}_B$. The attention weights provide a finer-grained reflection of the relationships between different check-in points within the sequences. Finally, we use the attention weight map from the last layer to guide our masking process.

\subsection{Masked Transformer Encoder} 
Inspired by the random masking strategy used in natural language processing~\cite{devlin2018bert, li2021mst}, we explored whether a masking strategy could be effectively applied to the User Identity Linkage (UIL) task. However, a random masking strategy could result in the loss of critical information within the check-in sequences ${\cal T}$, potentially impacting the model's overall performance. We propose an attention-guided masking module to preserve the positional correlations better and capture the co-occurrences within user check-in sequences. This module selectively masks non-essential tokens in the sequence, after which the masked sequence is encoded using a masked transformer encoder ${f_{\rm M}}(\cdot)$.

In the spatio-temporal embedding layer, we obtain the embeddings ${\bf{Z}}^A$, ${\bf{Z}}^B$ and ${\bf{Attn}}_{A \to B}$,  ${\bf{Attn}}_{B \to A}$ from the correlation attention block. Each token in the input embedding sequence is represented by vector sets $[{{\bf{z}}_1}, \ldots ,{{\bf{z}}_k}] \in {\bf{Z}}$. The attention weights along the key vector dimensions of ${\bf{Attn}}_{A \to B}$ and ${\bf{Attn}}_{B \to A}$ are summed as follows:
\begin{equation}
\begin{split}
N_{mask}^B,{{\bf{I}}^B} = {\mathop{\rm IdxValPair}\nolimits} (r, {\sum\limits_{j = 1}^{\left| {\bf{K}} \right|} {{{({\bf{Attn}}_{B \to A}^{\left| L \right|})}_{ij}}} } ),\\
N_{mask}^A,{{\bf{I}}^A} = {\mathop{\rm IdxValPair}\nolimits} (r, {\sum\limits_{j = 1}^{\left| {\bf{K}} \right|} {{{({\bf{Attn}}_{A \to B}^{\left| L \right|})}_{ij}}} } ),
\end{split}
\end{equation}
where the $\rm IdxValPair(\cdot)$ takes as input two components: a one-dimensional vector, which is the sum of the attention vector along the $j$-th dimension of the key vector $\bf{K}$ in the $i$-th batch, and the mask ratio $r$. It outputs the number of masked tokens $N_{mask}$ and their corresponding index vector ${\bf I} \in {{\mathbb R}^{ 1 \times {\left| {\bf{K}} \right|}}}$. 

Then top-k tokens with the lowest weights are selected as our candidate mask set. \ie, ${\rm set}_B$ and ${\rm set}_A$ .This strategy is represented as follows:
\begin{equation}
\left( {{\bf{m}} \odot {\bf{z}}} \right) = \left\{ {\begin{array}{*{20}{c}}
{{{\bf{z}}_{_\alpha }},  \quad    {m_i} = 1}.\\
{{{\bf{z}}_i},  \quad   \: {m_i} = 0}.
\end{array}} \right.
\end{equation}

Here, $\left( {{\bf{m}} \odot {\bf{z}}} \right)$ represents the final masked token, ${{\bf{z}}_i}$ is the vector at the $i$-th token position($i\le k$). ${{\bf{z}}_{_\alpha }}$ is a learnable mask embedding and $m_i$ is the $i$-th element of the ${\bf{m}} \in {{\bf{M}}_{mask}}$. If ${\bf{Att}}{{\bf{n}}_k}[i]$ is among the lowest $r$, $m_i$ is set to 1; otherwise, it is set to 0. ${\rm set}_A$ guides the masking for check-in sequence ${\cal T}_A$, and ${\rm set}_B$ for check-in sequence ${\cal T}_B$. The masked embeddings for platforms $A$ and $B$, \ie, ${\bf{Z}}_A^m$ and ${\bf{Z}}_B^m$ are obtained through the $Hadamard$ product and matrix addition. The final representation ${\bf{H}}_A^m$ and ${\bf{H}}_B^m$ is derived by encoding these masked embeddings with ${f_{{\rm M}}}(\cdot)$.

\subsection{User Identity Linkage Layer}
Our user linkage layer draws inspiration from ~\cite{chen2024trajectory}, using a multi-layer feed-forward neural network as the predictor. We employ a $sigmoid$ function as the logistic regression function to generate the final similarity score. The sequence representations ${\bf{H}}_A^m$ and ${\bf{H}}_B^m$ are concatenated to form the final feature representation, which is then fed into the predictor to obtain the similarity score. This process can be viewed as a binary classification task, where Eq.~\eqref{eq.one} determines whether the two check-in sequences belong to the same user. 

We optimize the final normalized probability using a binary cross-entropy loss function. The binary cross-entropy loss for a single sample is given by:
\begin{equation}
\begin{split}
\label{eq.eight}
{{\cal L}}(y,\hat y) =  - \frac{1}{N}\sum\limits_{i = 1}^N {{\bf{w}}_i}  \cdot \ell ({y_i},{\hat y_i}),
\end{split}
\end{equation}
where $\ell ({y_i},{\hat y_i}) = {y_i}\log ({{\hat y}_i}) + (1 - {y_i})\log (1 - {{\hat y}_i})$ and $y$ represents whether $u^{A}_i$ and $u^{B}_j$ are the same users, represented by 0 and 1; $\hat y$ represents the classification result of the model. $N$ represents the total number of samples and ${{\bf{w}}_i}$ represents the corresponding weighted item (selected according to the label).
The entire process can be referenced in Algorithm \ref{alg:MTE}.


\begin{algorithm}[tb]
  \caption{The Learning Process of \model}
  \label{alg:MTE}
  \begin{algorithmic}[1]
  \renewcommand{\algorithmicrequire}{\textbf{Input}}
  \renewcommand{\algorithmicensure}{\textbf{Output}}
    \REQUIRE
      User spatial-temporal check-in sequences ${\cal T}_A$, ${\cal T}_B$ and timestamp $t_A$ and $t_B$
    \ENSURE
      Prediction probability $p$

    \STATE Get ${\bf X}_{A/B} = f_{st}({\cal T}_{A/B})$;
    \STATE Get ${\bf Z}_{A/B} = {\bf X}_{A/B} + {\rm TE}(t_{A/B})$;
    \STATE Get ${\bf H}_{A/B} = {f_{\rm T}}({\bf Z}_{A/B})$;
    \FOR {${l}:{|\mathit{L}|}$}
    \STATE Get ${{\bf Q}}$, ${{\bf K}}$, ${{\bf V}}$ from $\rm linear({\bf H}_{A/B})$;
    \STATE ${\bf{Attn}}_{{A \to B}/{A \to B}}$ = ${\rm CoAttn}({\bf Q}_{A/B}, {\bf K}_{B/A}, {\bf V}_{B/A})$;
    \ENDFOR
    \STATE Initialize ${{\bf{M}}_{mask}}$ as zero matrix ${\bf{0}}$;
    \STATE Get $N^{B/A}_{mask}$ from ${\bf{Attn}}_{{A \to B}/{B \to A}}$;
    \FOR{${k}:{N^{B/A}_{mask}}$}
    \STATE ${{\bf{M}}_{mask}}[k]$=${\mathop{\rm topk}\nolimits} ({\mathop{\rm sort}\nolimits} ({\bf{Attn}}_{{A \to B}/{B \to A}}^{\left[ k\right]}))$;
    \ENDFOR
    \STATE ${{\bf{Z}}^m_{B/A}} = (1 - {{\bf{M}}_{mask}}) \odot {{\bf{Z}}_{B/A}} + {{\bf{M}}_{mask}} \odot {{\bf{z}}_{\alpha}}$;
    \STATE ${{\bf{H}}^m_{B/A}} = {f_{{\rm M}}}({{\bf{Z}}^m_{B/A}})$;
    \STATE $y = \sigma ([{{\bf{H}}^m_A};{{\bf{H}}^m_B}])$;
    \STATE Calculate $\mathcal{L}$ with Eq.~\eqref{eq.eight} through $\hat y$ and $y$;
    \STATE Back propagation and update parameters in \model;
    \RETURN $p$
\end{algorithmic}
\end{algorithm}

\begin{table}[ht]
\centering
\setlength{\tabcolsep}{1.8mm} 
\renewcommand{\arraystretch}{1} 

\begin{tabular}{llccc}
\toprule
{Datasets} & {Platforms} & {\#Users} & {\#Records} & {\#Trajs} \\
\midrule                   
\multirow{3}{*}{{XSiteTraj}} & {Twitter}     & 11,239 & 1,187,063 & 11,239 \\
                                 & {Foursquare}  & 8,569  & 232,932  & 8,569  \\
                                 & {Facebook}    & 7,146  & 309,664  & 7,146  \\
\midrule
\multirow{2}{*}{\makecell{{ISP} \\ {-Weibo}}} & {ISP}     & 30,405 & 3,423,865 & 202,656 \\
                                                          & {Weibo}   & 20,624 & 175,895  & 20,624  \\
\bottomrule
\end{tabular}
\caption{Statistics of the datasets.}
\label{tab:dataset_statistics}
\end{table}

\section{Experiments}
\begin{table*}[ht]
\centering
\fontsize{9}{12}\selectfont
\setlength{\tabcolsep}{0.2mm} 
\renewcommand{\arraystretch}{1.1} 

\begin{tabular}{c|cccc|cccc|cccc|cccc}
\toprule
\textbf{} & \multicolumn{4}{c|}{Twitter-Foursquare} & \multicolumn{4}{c|}{Twitter-Facebook} & \multicolumn{4}{c|}{Foursquare-Facebook} & \multicolumn{4}{c}{ISP-Weibo} \\
\cline{2-17}
\textbf{} & P(\%) & R(\%) & F1(\%) & 
AUC(\%) & P(\%) & R(\%) & F1(\%) & AUC(\%) & P(\%) & R(\%) & F1(\%) & AUC(\%) & P(\%) & R(\%) & F1(\%) & AUC(\%) \\
\midrule
DPLink     & 51.40 & 52.38 & 51.08 & 51.91 & 64.52 & 54.97 & 55.39 & 52.29 & 53.34 & 51.83 & 51.88 & 53.78 & 24.99 & 50.00 & 33.32 & 50.57\\
DPLink-SM  & 55.42 & 53.02 & 52.17 & 52.27 & \underline{66.69} & 55.40 & 55.99 & 53.21 & \underline{54.25} & 51.49 & 51.02 & 55.32 & 50.70 & 53.09 & 40.64 & 52.88 \\
CPLink     & 53.32 & 58.57 & 55.80 & 74.97 & 60.34 & 61.19 & 60.74 & 77.21 & 46.41 & \underline{62.90} & 53.40 & \underline{79.53} & \underline{63.07} & \underline{70.72} & \underline{66.67} & \underline{80.17}\\
CPLink+    & \underline{58.76} & \underline{65.67} & \underline{62.01} & \underline{78.96} & 63.06 & \underline{68.04} & \underline{65.44} & \underline{80.67} & 47.95 & 62.28 & \underline{54.10} & 79.33 & 63.01 & 69.34 & 66.02 & 79.57\\
\textbf{Ours}        & \textbf{66.14$^*$} & \textbf{73.21$^*$} & \textbf{68.06$^*$} & \textbf{82.43$^*$} & \textbf{71.09$^*$} & \textbf{80.46$^*$} & \textbf{73.82$^*$} & \textbf{87.86$^*$} & \textbf{57.34$^*$} & \textbf{72.94$^*$} & \textbf{62.87$^*$} & \textbf{82.80$^*$} & \textbf{80.37$^*$} & \textbf{79.46$^*$} & \textbf{79.85$^*$} & \textbf{86.89$^*$} \\
\cline{1-17}
\textit{Impro.}    & \textbf{12.55} & \textbf{11.48} & \textbf{9.75} & \textbf{4.39} & \textbf{6.59} & \textbf{18.25} & \textbf{12.80} & \textbf{8.91} & \textbf{5.69} & \textbf{15.96} & \textbf{16.21} & \textbf{4.11} & \textbf{27.42} & \textbf{12.35} & \textbf{19.76} & \textbf{8.38} \\
\bottomrule
\end{tabular}
\caption{Performance comparison of all models on four real-world datasets. P represents Macro-Precision, R represents Macro-Recall, and F1 represents Macro-F1. Marker * indicates the
results are statistically significant (t-test with p-value $<$ 0.01).}
\label{tab:exp_result}
\end{table*}

In this section, we evaluate our proposed model using four real-world cross-platform check-in datasets.
\subsection{Datasets}
We collect two publicly available cross-platform check-in datasets: (1) XSiteTraj~\cite{fu2023xsitetraj} and (2) ISP-Weibo~\cite{feng2019dplink}. The XSiteTraj dataset includes data from three platforms: Twitter, Foursquare, and Facebook, which allowed us to create three cross-platform datasets: Twitter-Foursquare, Twitter-Facebook, and Foursquare-Facebook. Due to differences in the datasets and significant variations in trajectory lengths, we apply different preprocessing steps to each dataset. To facilitate training, we remove sequences longer than 400, 200, and 200 from the Twitter, Foursquare, and Facebook datasets, respectively, as these represent a small proportion of the data (5.20\%, 0.28\%, 0.17\%). We also remove sequences shorter than 3 from the ISP and Weibo datasets, as they account for a small percentage of the data (3.53\%, 0.13\%). Table~\ref{tab:dataset_statistics} presents the statistical details of the preprocessed datasets.

\subsection{Baselines}


\begin{itemize}
\item \textbf{DPLink}~\cite{feng2019dplink}: A deep learning model based on RNN that introduces an end-to-end deep learning framework to extract spatio-temporal locality features for user linkage.

\item \textbf{DPLink-SM}~\cite{feng2020user}: An extended version of DPLink, which incorporates a trajectory similarity matcher to aid in the user identity linkage task, improving on DPLink's capabilities.

\item \textbf{CPLink}~\cite{ding2020user}: A data mining approach that addresses the spatio-temporal UIL task using frequent pattern mining and clustering techniques.

\item \textbf{CPLink+}~\cite{ma2022cp}: An enhanced version of CPLink that replaces the binary value function with a real-valued function to calculate the overlap area and uses the original DTW (Dynamic Time Warping) algorithm to achieve optimal distance measurement.
\end{itemize}

\subsection{Evaluation Metrics and Experiment Settings}
In our evaluation, we use Macro-Precision, Macro-Recall, Macro-F1, and Area Under the Curve (AUC) to quantify the performance of different methods. 

For the baseline models, we use the parameter settings recommended in their respective papers and fine-tune them for optimal performance. The check-in embedding dimension for our model is set to 64, the mask ratio $r$ to 0.1, and the initial learning rate to 0.001. The dropout rate is set to 0.1, and the correlation attention block consists of 2 layers. We apply early stopping during validation with a patience of 5 to prevent overfitting. Each experiment is repeated five times, and the average results for all methods are reported. All experiments are conducted on a machine with NVIDIA GeForce RTX 3090 GPU.

\subsection{Experimental Results}

Table~\ref{tab:exp_result} summarizes our experimental results, with the best outcomes highlighted in \textbf{bold} and the second-best \underline{underlined}. Our model aims to address two core challenges in the current spatio-temporal UIL task: 1) \textit{retaining co-occurrence points while filtering out noise}, and 2) \textit{capturing the spatio-temporal co-occurrence between user check-in sequences across different platforms.} Our experimental results validate the effectiveness of \model in tackling these challenges, significantly outperforming existing comparison methods.

First, in the Twitter-Foursquare, Foursquare-Facebook, and Twitter-Facebook datasets, our model excelled in precision and recall, with average improvements of 11.48\%, 15.96\% and 18.25\%, respectively. This demonstrates that \model successfully captures the spatio-temporal co-occurrence information between cross-platform users while effectively preserving key spatio-temporal points. By ignoring noise points that could negatively impact the model, \model focuses on the most relevant features. The spatio-temporal co-linearity information and its relational features effectively guide the masking process, enabling \model to learn robust sequence representations. DPLink and DPLink-SM fail to account for the spatio-temporal co-occurrence of key points between cross-platform users, leading to lower classification performance. On the other hand, CPLink and CPLink+ are limited by larger datasets, which restrict their ability to effectively capture co-occurring key points. This makes it difficult for these models to accurately identify critical spatio-temporal patterns, further impacting their performance in complex cross-platform scenarios.

Second, in ISP-Weibo datasets, our model shows a significant advantage in Precision, with a significant improvement of 27.42\%. Our model has demonstrated its effectiveness by maintaining high performance even on large datasets. This indicates that the model successfully captures spatio-temporal co-occurrence between cross-platform users. In contrast, DPLink and DPLink-SM struggle with large datasets and have difficulty maintaining classification accuracy in noisy environments. The lower accuracy of CPLink and CPLink+ on the current datasets suggests that data mining methods are limited by manually constructed features and frequent pattern mining, preventing the models from autonomously learning abstract semantic representations. Furthermore, CPLink and CPLink+ do not leverage real labels to distinguish users, making them more susceptible to false samples and leading to incorrect user identity recognition.

\begin{table}[]
\centering
\resizebox{8.5cm}{!}{ 
\begin{tabular}{ccccc}
\hline
\multirow{2}{*}{Method} & \multicolumn{2}{c}{Twitter-Foursquare} & \multicolumn{2}{c}{Twitter-Facebook} \\ \cline{2-3}  \cline{4-5} 
                        & F1                 & AUC               & F1            & AUC           \\ \hline
\textit{w/o} MTE               & 62.05              & 64.33             & 71.63            & 83.33            \\
\textit{w/o} CAB             & 56.17              & 67.04             & 67.10            & 82.20            \\
\textit{w/o} TTE                  & 54.20              & 58.38                & 68.52            & 84.98            \\ \hline
MT-Link                 & 68.06              & 82.43             & 73.82            & 87.86            \\ \hline
\end{tabular}
}
\caption{The results of ablation study.}
\label{tab:Ablation}
\end{table}

\begin{figure*}[h]
    \begin{center}
    \resizebox{\textwidth}{!}{%
        \subfigure[Mask ratio $r$ on Tw-Fb.]{
        \label{fig:a}
        \includegraphics[height=0.15\textwidth]{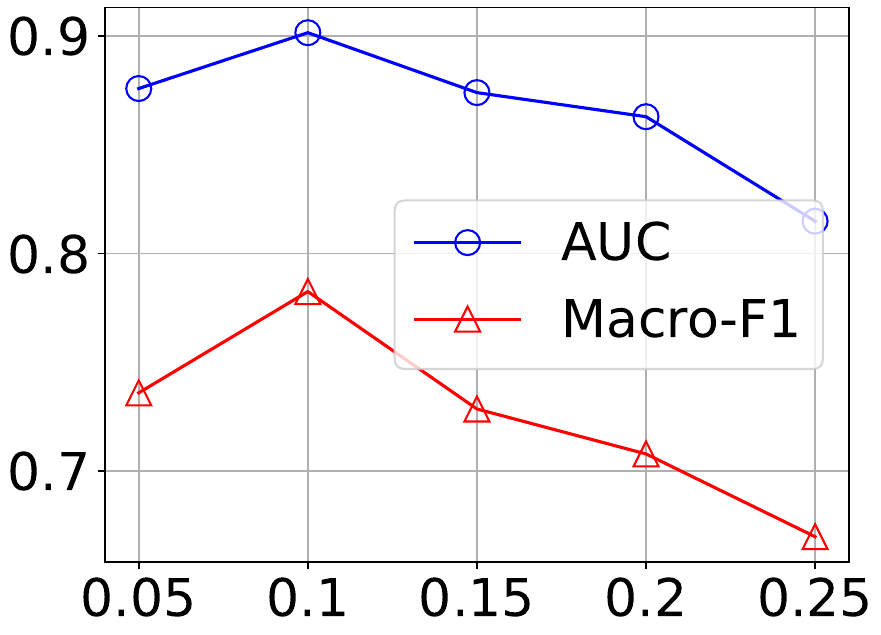}
        }
        \subfigure[Mask ratio $r$ on Tw-Fs.]{
        \label{fig:b}
        \includegraphics[height=0.15\textwidth]{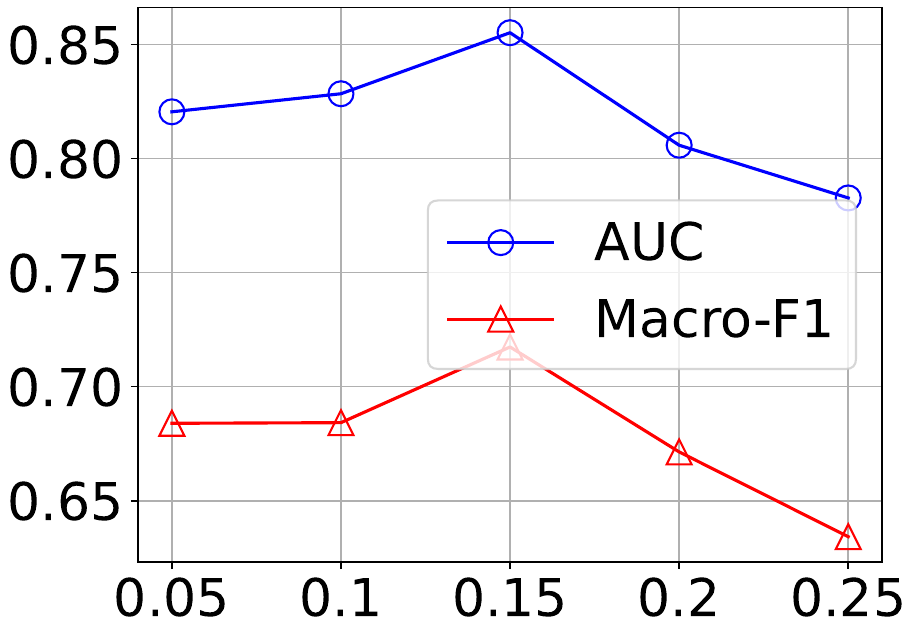}
        }
        \subfigure[\#layers on Tw-Fb.]{
        \label{fig:c}
        \includegraphics[height=0.15\textwidth]{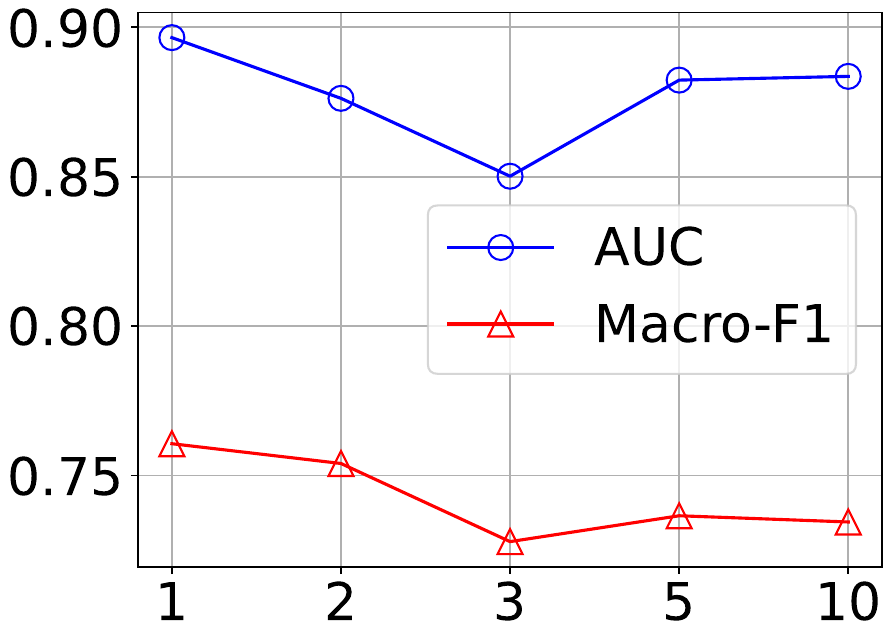}
        }
        \subfigure[\#layers on Tw-Fs.]{
        \label{fig:d}
        \includegraphics[height=0.15\textwidth]{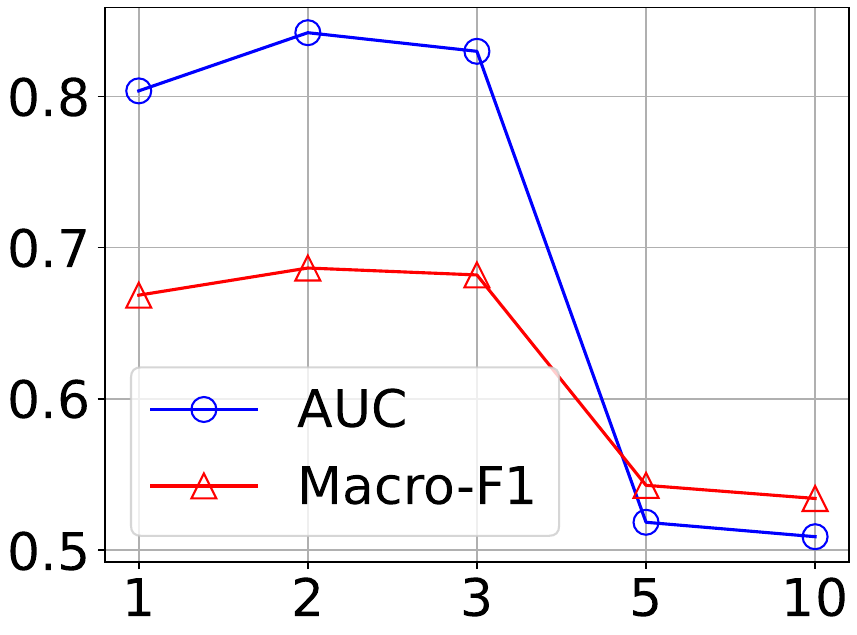}
        }
    }
    \vspace{-2.99mm}
    \caption{The impact of mask ratio and correlation attention block layers on \model. Tw-Fb represents the Twitter-Facebook dataset, and Tw-Fs represents the Twitter-Foursquare dataset.}
    \label{fig:sensitive}
    \vspace{-5mm}
    \end{center}
\end{figure*}

\subsubsection{Ablation Study.}
To verify the effectiveness of key components, we conduct ablation experiments (as shown in Figure~\ref{tab:Ablation}) by removing components of \model. The obtained model variants are as follows:
\begin{itemize}

    \item \textbf{\textit{w/o} MTE}: We remove the masked transformer encoder from the model and use the output of the correlation attention block directly for the classification task.
    
    \item \textbf{\textit{w/o} CAB}: We remove the correlation attention block from the model and replace it with a random masking strategy to guide the masking process.
    
    \item \textbf{\textit{w/o} TTE}: We remove the temporal transformer encoder and replace it with a standard transformer encoder.

\end{itemize}
As seen in Table~\ref{tab:Ablation}, \model contributes effectively to its overall performance.
\textbf{\textit{w/o} MTE} compares with our \model, it is evident that masked transformer effectively preserves key spatio-temporal co-occurrence points while filtering out noise points. Compared to the \textit{w/o} MTE, \model significantly increases AUC by 23.91\% on Twitter-Foursquare. This approach strengthens the model's ability to effectively filters out noise, leading to improved performance in user identity linkage tasks. \textbf{\textit{w/o} CAB} suggests that the correlation attention block plays a critical role in guiding the retention of key points (those with high attention weights) and in capturing spatio-temporal co-occurrence between user check-in sequences on different platforms. \textbf{\textit{w/o} TTE} shows that temporal position encoding significantly impacts datasets with long sequences, particularly in the Twitter-Foursquare dataset, where AUC significantly decreased by 41.19\% when \textit{w/o} TTE. This indicates that the temporal position encoder is crucial for generating spatio-temporal embeddings of check-in sequences.

\subsubsection{Parameter Sensitivity.}

We also evaluate the sensitivity of \model to different mask ratios and correlation attention block layers. Figure~\ref{fig:sensitive} shows the performance of masking rate and correlation attention block layers on Twitter-Facebook and Twitter-Foursquare. 

We find that the performance on both datasets initially increases with the rise in the mask rate, reaches a critical value, and then begins to decline. The optimal mask rates are 0.1 and 0.15, which ensure that most key information is retained while effectively removing noise, meeting the model's requirements.

In addition, the performance on Twitter-Facebook remains relatively stable across different numbers of correlation attention block layers, with optimal performance at one layer. However, on Twitter-Foursquare, performance drops sharply when the number of layers is increased, suggesting that too many layers can disrupt key features, preventing accurate reflection of cross-platform user check-in co-occurrence.

\begin{figure}[h]
    
    \begin{minipage}[t]{0.32\linewidth} 
        \subfigure[${\bf{Attn}}_{A \to B}$.]{
        \label{fig:b}
        \includegraphics[trim=31 35 0 0, clip, width=\linewidth]{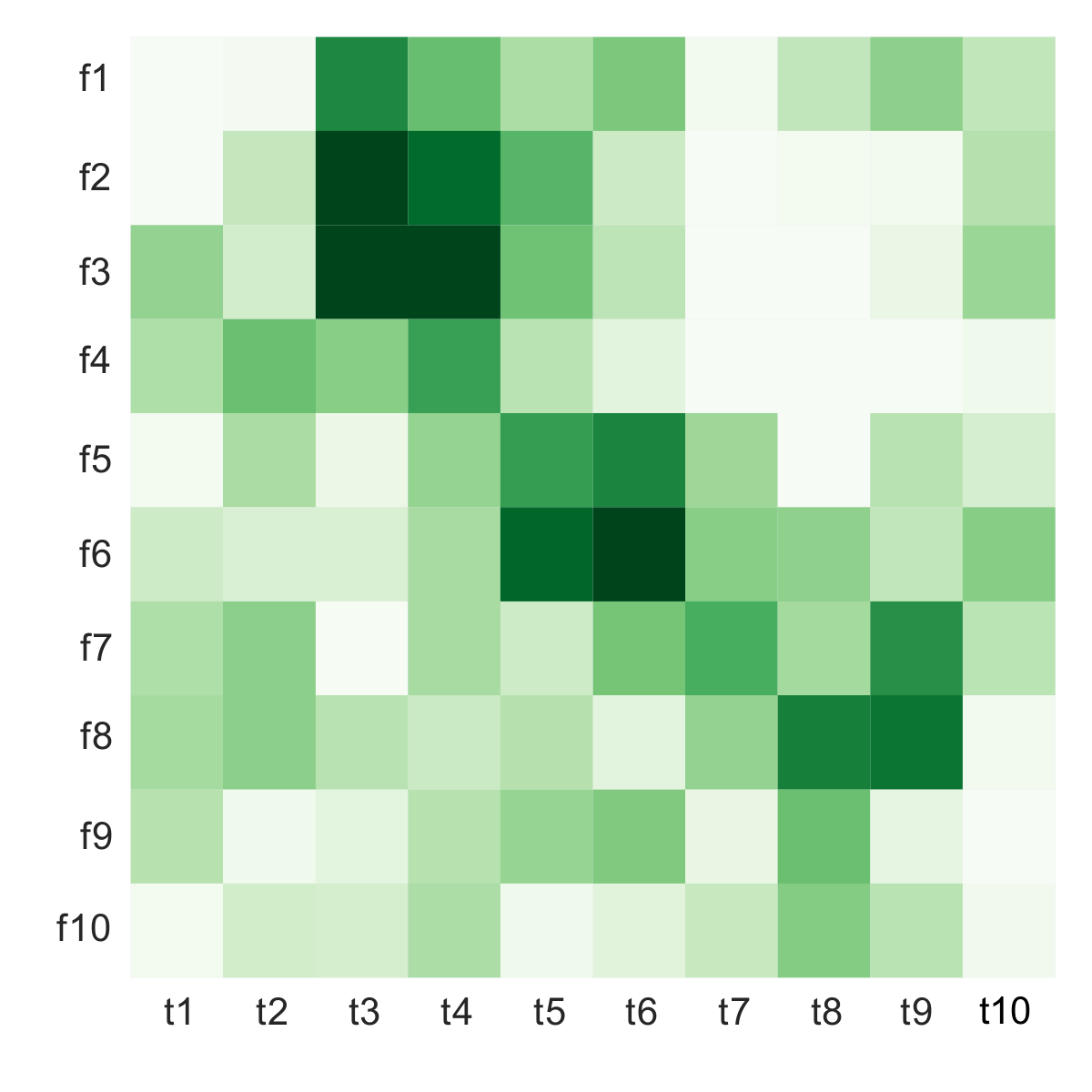} 
        }
    \end{minipage}%
    \hspace{0.001\linewidth} 
    \begin{minipage}[t]{0.32\linewidth}
        \subfigure[${\bf{Attn}}_{B \to A}$.]{
        \label{fig:c}
        \includegraphics[trim=31 35 0 0, clip, width=\linewidth]{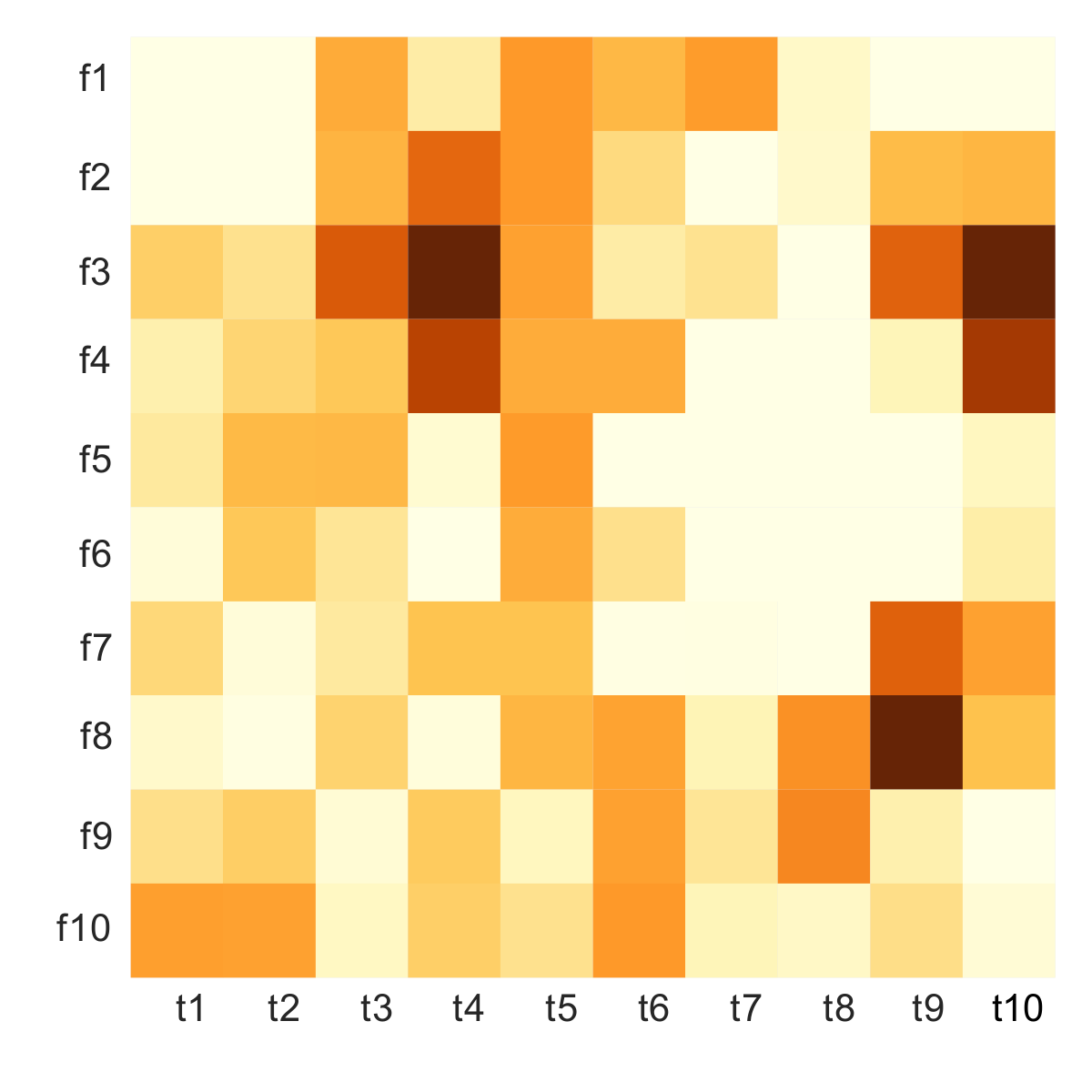}
        }
    \end{minipage}
    \hspace{0.001\linewidth} 
    \begin{minipage}[t]{0.32\linewidth}
        \subfigure[Co-occurrences.]{
        \label{fig:d}
        \includegraphics[trim=31 35 0 0, clip, width=\linewidth]{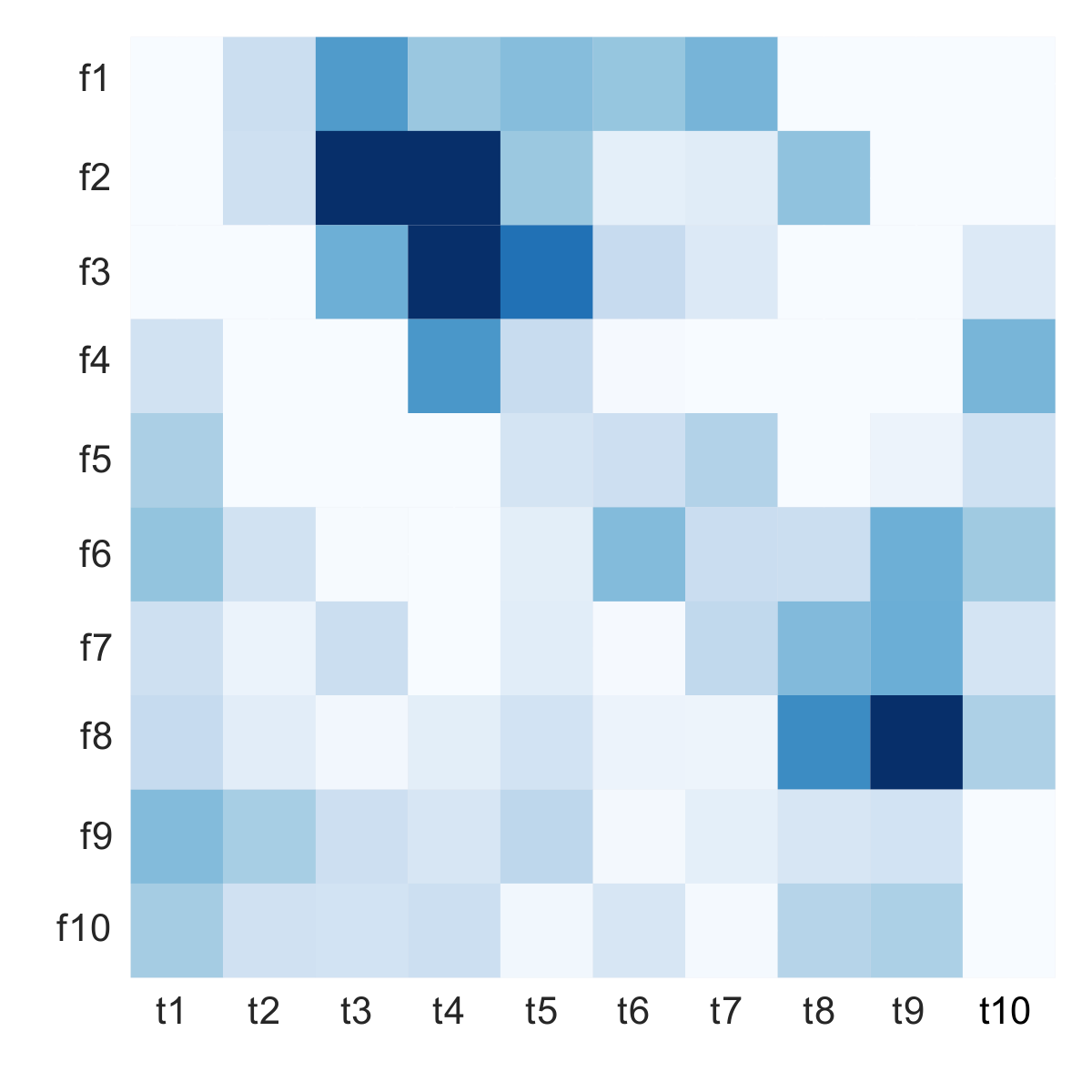}
        }
    \end{minipage}

    \vspace{-3mm}
    \caption{Co-occurrence visualization.}
    \label{fig:Verification}
    \vspace{-5mm}
\end{figure}

\subsubsection{Co-occurrence Visualization.}
We output the attention weight maps learned from the correlation attention block for two trajectories, each with ten check-ins from Twitter and Foursquare, denoted as ${\bf{Attn}}_{A \to B}$ and ${\bf{Attn}}_{B \to A}$, as shown in Figure~\ref{fig:Verification}(a) and Figure~\ref{fig:Verification}(b) (\ie, with the $x$-axis representing Twitter check-ins and the $y$-axis representing Foursquare check-ins). We also calculate the pairwise distances between check-in locations of two trajectories and normalize them to obtain a co-occurrence matrix, as shown in Figure~\ref{fig:Verification}(c). Higher values indicate stronger spatio-temporal co-occurrences between the corresponding check-in points (darker colors), while lower values indicate weaker co-occurrences (lighter colors). 
By examining Figure~\ref{fig:Verification}(a) and Figure~\ref{fig:Verification}(b) alongside the ground truth Figure~\ref{fig:Verification}(c), we can conclude that \model correctly learn spatio-temporal co-occurrence patterns at these check-in points, for example, \textit{Grid}($t9, f8$) and \textit{Region}($t3~f2$, $t4~f3$) exhibit similar strong spatio-temporal co-occurrence as in Figure~\ref{fig:Verification}(c), and \textit{Grid}($t10, f9$) and \textit{Region}($t8~f3$, $t9~f5$) also show similar weaker co-occurrence as in Figure~\ref{fig:Verification}(c). 
However, \textit{Grid}($t6, f6$) in Figure~\ref{fig:Verification}(a) and \textit{Grid}($t10, f3$) in Figure~\ref{fig:Verification}(b) show stronger co-occurrence than their counterparts in Figure~\ref{fig:Verification}(c), suggesting that \model identifies important co-occurring check-in points when modeling correlations across different trajectories. In conclusion, the two learned attention matrices exhibit regional weights similar to those in the co-occurrence matrix, demonstrating that our model effectively captures accurate spatio-temporal co-occurrence information across user check-in sequences. 

\subsubsection{Computation Time.}

\begin{figure}
    \begin{center}
    \includegraphics[width=0.43\textwidth]{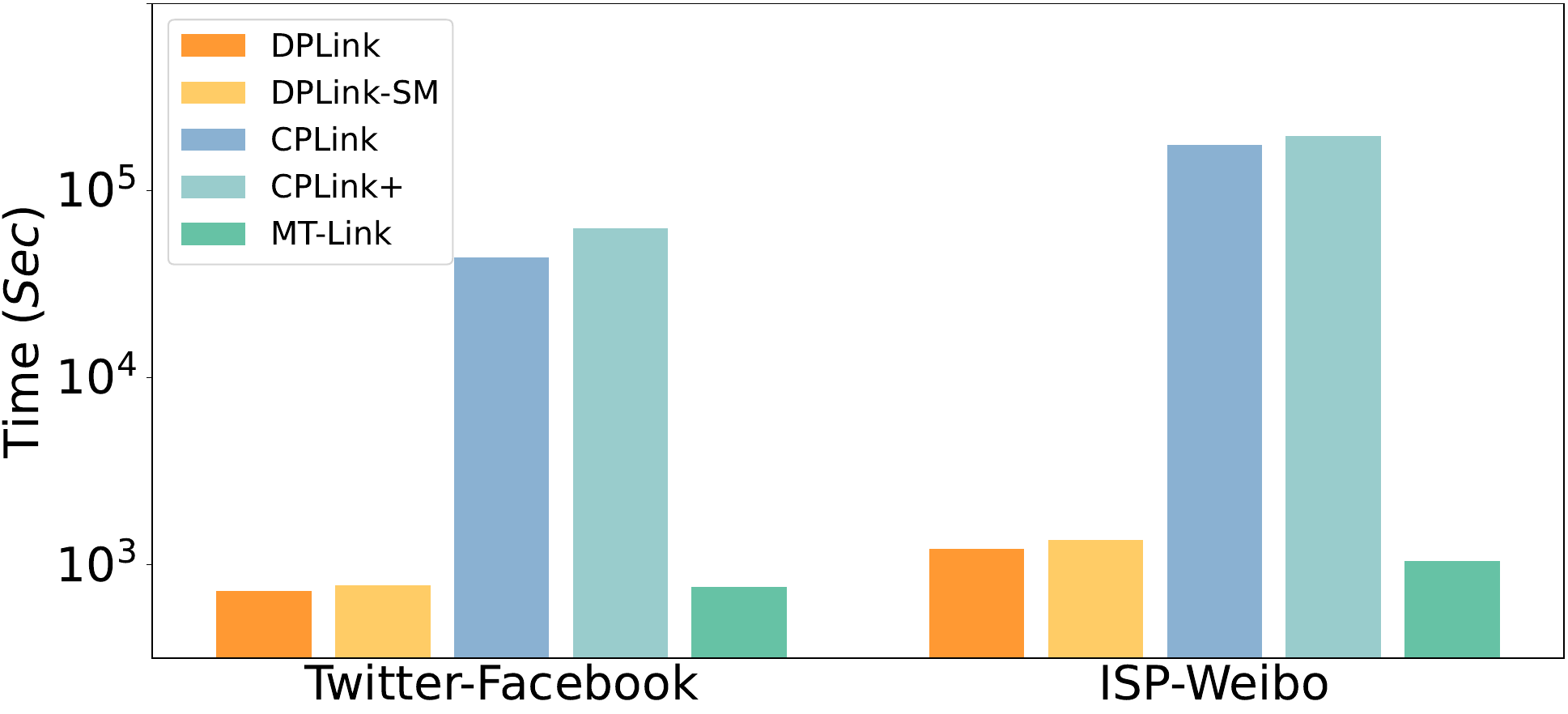}
    \vspace{-1mm}
    \caption{The time costs of \model and other baselines.}
    \vspace{-5mm}
    \label{fig:time}
    \end{center}
\end{figure}

To validate the effectiveness of \model in addressing the third challenge, we statistically analyze the runtime of all baselines. As shown in Figure~\ref{fig:time}, CPLink's reliance on traditional data mining and clustering tasks results in high time complexity. Specifically, when applied to ISP-Weibo dataset, the running time for traditional data mining methods dramatically increases, about 187 times that of \model, highlighting their limitations on large-scale datasets. 
Compared with DPLink and and its successor DPLink-SM, \model consumes similar running time. This may be because although our \model does not require pre-training, it uses masking strategy and correlation attention block, which requires additional time to mine key information. Despite this, \model has significantly better performance than DPLink and DPLink-SM.


\section{Conclusion}

In this paper, we propose a transformer-based framework, \model, to enhance model performance by learning spatio-temporal co-occurrence patterns of cross-platform users. Our model captures the spatio-temporal co-occurrence between users and utilizes a masked transformer to filter out noise points. In extensive experiments on four real-world cross-platform datasets, our model significantly outperforms state-of-the-art baselines.



\section*{Acknowledgments}
This work is partially supported by the National Natural Science Foundation of China under Grant No. 62176243, the Fundamental Research Funds for the Central Universities under Grant No 202442005, and the National Key R\&D Program of China under Grant No 2022ZD0117201. 


\bibliography{cite}



\end{document}